# Reviewing Climate Change and Agricultural Market Competitiveness

Bakhtmina Zia[1]

PhD Research Scholar, Institute of Management Sciences, Peshawar, Pakistan

Dr Muhammad Rafiq[2]

Associate Professor, Institute of Management Sciences, Peshawar, Pakistan

**Abstract**

The paper is a collection of knowledge regarding the phenomenon of climate change, competitiveness; and literature linking the two phenomena to agricultural market competitiveness. The objective is to investigate the peer-reviewed and grey literature on the subject to explore the link between climate change and agricultural market competitiveness and also explore an appropriate technique to test/validate the presumed relationship empirically. The paper concludes by identifying implications for developing an agricultural competitiveness index while incorporating the climate change impacts; to enhance the potential of agricultural markets for optimizing the agricultural sectors' competitiveness.

**1. Bakhtmina Zia, PhD Research Scholar, Institute of Management Sciences, I-A, Sector E-5, Phase VII, Hayatabad, Peshawar, Pakistan**

+92-321-9186906, +92-317-8510226

bakhtmina.zia@gmail.com ; phd174106363@imsciences.edu.pk

# 1.0 Introduction

The United Nations Framework Convention on Climate Change described climate change in 1992 as "*a change of climate which is attributed directly to human activity that alters the composition of the global atmosphere and which is in addition to natural climate variability observed over comparable periods*". The change is an outcome of many activities including temperature changes, precipitation changes, sunlight hours, and sun energy changes. Further, the combusting fuels generating greenhouse gases, modernization, and urbanization process accompanied by the industrial growth and deforestation also contribute to climate change (Upreti, 1998). These result in perpetual changes in all spheres of human life, bringing hazardous outcomes in areas including natural resources, habitats, geological processes, agriculture, forests, and vegetation most of which pose serious threats to human life and food security (Malla, 2008). Like other areas, the implications can be serious for market competitiveness as well.

The market competitiveness as defined by the Global Competitiveness Index 2015-16 is a composite index ascertaining the country's productivity levels, while productivity level serves as the determinant of the country's prosperity level. Similarly, the competitiveness of any sector determines the prosperity and standing of that sector in an economy. Gardiner et al. (2004) state that only nations approaching high productivity indexes and record-high employment rates can be competitive. While Laureti and Viviani (2011), established that firm competitiveness considerably gets effected by productivity. These outcomes endorse Porter's theory suggesting that firm competitiveness and productivity are analytically linked up, making them helpful in improved decision making for increasing competitiveness at the firm level. Whilst several scholars stress that national productivity is the sole representative phenomenon for competitiveness measurement (Porter, 1990). Just like the firms, the farm (agricultural) productivity is an important determinant of agricultural competitiveness. However, the literature from Krugman (1994) and others declare productivity to be a good measure of competitiveness only at the national level, and not at the sectoral or firm level.

The impact of climate change has been studied by many researchers as can be seen indifferent studies (Kumar, A., & Sharma, P., 2013; Baig.et.al, 2011; Gornall, J.et.al, 2010; Shakoor et.al, 2011; Mendelson et.al. , 1992; Rehana et.al, 2012). The climate change impacts on agricultural farm revenues, wheat production, and other major crops were studied in the above studies by using different climate variables like rainfall, temperature, etc. Most of these studies concluded with the fact that climate change had a measurable impact on farm productivity and revenue. Studies are also available measuring climate change impact on food security, adaptation, and mitigation policies (Wheeler, T., & Braun, J., 2013). According to the IPCC Climate Change report 2014, "it is clear from the literature that the changing climate has more negative impacts as compared to the positive impacts in a large number of crops and regions. While a small number of studies, revealing the positive impacts of changing climate are mostly targeting the regions with high latitudes, and still, it is ambiguous whether the overall balance of impacts in the region are still positive or become negative. Moreover, the market access, availability, and other market-related components of food security still appear to be neglected as most of the existing studies target only the production side of food security. However, the existing markets in important producing regions, seem to be highly responsive to stages of quick price hikes resulting from the climate extremes". Thus, the changing climate's impact on agricultural market competitiveness seems to be ignored to date and the area needs to be explored to find the existent relationship.

Against this background, this literature review paper addresses the question: what is the relation between climate change and agricultural market competitiveness, and what are the important factors which determine this relation? This paper investigates the peer-reviewed and grey literature on the subject to explore the link between climate change and agricultural market competitiveness and also explores an appropriate technique to test/validate the presumed relationship empirically. This will help in designing agricultural market-friendly climate policies and thus, the local, as well as international competitiveness of the agriculture sector, may be enhanced.

**2.0 Climate Change**

The Australian Academy of Science declares climate change as a changing weather pattern with subsequent longer period changes in the icebergs, earth surfaces, and oceans. Additionally, it is the variation in demographic properties of climate system persistent over decades with a minimum of 30 years, having the extremes, variability, and averages serving as its core properties. The changing climate is an outcome either of natural processes including changing sun radiations, internal variations in the climate system or the volcanoes; or of human-induced changes affecting the land use and atmospheric structure. Though the climate has changed several times in the earth's history; a significant issue with the current change in climate is that human beings; for the most part are responsible for it, leading to higher levels of greenhouse gas emissions. Such intrusions in the climate system result in different unprecedented changes in the climate system; causing rising temperatures and sea levels, melting of ice and snow, frequent floods, and droughts along with other serious impacts (Environmental Protection Agency).

The United States National Aeronautics and Space Administration (NASA) describes climate change as "a long-term change in the average weather patterns that have come to define Earth's local, regional and global climates. These changes have a broad range of observed effects that are synonymous with the term".

**2.1 Measuring Climate Change**

The basic elements that are commonly used for measuring changes in climate are mainly classified in; temperature, precipitation, biomass, sea level, Solar activity, volcanic eruptions, & chemical composition. (World Meteorological Organization (WMO), 2018; IPCC, 2007 Fourth Report). However, the National Oceanic Atmospheric Administration (NOAA), enumerates Essential Climate Variables (ECVs), to find about the changing climatic conditions. They categorize the variables into three broad groups as terrestrial (land), atmospheric (air), and oceanic (ocean). Among these, the Terrestrial Climate Variables include river discharge, glaciers, and ice caps, ice sheets, permafrost, albedo, land cover, soil carbon, fire disturbance, and soil moisture. While the atmospheric/air climate variables are further divided in to surface air variables including air temperature, precipitation, solar radiation, air pressure, wind, water vapor; and upper air variables including cloud properties and composition.

## 3.0 Competitiveness

### 3.1 Defining Competitiveness

Despite its existence, for many years the term competitiveness has remained an occult phenomenon. Although it is an important concept yet it lacks a universal definition. A reason for this is that competitiveness is a complex, multi-component, and a relative phenomenon. It is related to a large number of mutually dependent variables, which makes it hard to define this term. According to Webster's English Dictionary, the word 'competitiveness' originates from the Latin word 'competer', which means 'involvement in a business rivalry for markets'. The term in business parlance, however, generally means 'the ability to compete'. While the Cambridge dictionary explains competitiveness as "the ability of people to compete successfully".

A global level definition stated in The Global Competitiveness Report; defines competitiveness "as the set of institutions, policies, and factors that determine the level of productivity of a country." While productivity level serves as the determinant of the country's prosperity level. (WEF, 2014). On the other hand, the International Institute of Management Development World Competitiveness Yearbook declares competitiveness as, "the ability of a nation to create and maintain an environment that sustains more value creation for its enterprises and more prosperity for its people" (IMD World Competitiveness Center, 2014).

The work of global organizations mostly relies on productivity as an important representative of competitiveness. However, recent research enumerates other aspects of competitiveness as well. Atkinson (2013) discusses some of the popular definitions of competitiveness and argues that although productivity is considered synonymous with the concept of competitiveness, it has a quite different application in the practical world and cannot be a sole representative of the competitiveness phenomenon. Moreover, innovation is also linked to the phenomenon of competitiveness as Atkinson (2013) declares competitiveness to be related to productivity and innovation but still a distinct term from them, and therefore requires a different definition. Therefore, he goes on to define competitiveness as, "the capability of a region to export more in value-added terms than it imports." Also, he declares an economy to be competitive if it is having a trade surplus, not too many barriers to imports, and operates with restricted "discounts" to exporters.

Further, in qualitative content analysis, Plumins et al., (2016) made an exhaustive study of 125 competitiveness definitions for the period of 1949-2013 and after analysis of 34 definitions, they clearly stated competitiveness to be an ability to attain something through commodities on a global scale. They further explained the term under three-time categorizations of its constructs during its evolution. The early definitions of competitiveness were focused more upon the scale and core meaning of the term competitiveness. Afterward, the important channels especially the strategies and resources relating the commodities; for attaining and sustaining competitiveness became a locus of the definitions. Finally, internal factors including capabilities, skills, and competencies of organizations were included in the definitions. Besides, Plumins et al. (2016) found that competitiveness term was equally linked to the national and firm-level, while there was comparatively less discussion upon the sector, industry, and regional level.

Moreover, Lee & Karpova (2018) researched advancing the competitiveness theory-building process. Three major theories and the existing literature provided the basis for their new definition, with special emphasis on the current global environment. They enumerated three basic components to define the term competitiveness after conducting a systematic content analysis of nine existing definitions. Fundamental goals, procedures, and historical factors were the identified constructs. The principle goal was conceptualized as a better living standard based upon per capita incomes, employment, and equal income. In the methodological factor, productivity is shown as a fundamental source for attaining competitiveness. While in the background factor, knowledge is shown as the most important factor in determining competitiveness in the new world environment. Based on the identified factors, competitiveness was defined as,

 "Competitiveness is an ability to achieve a high standard of living through productivity growth in the new global environment, where knowledge becomes a critical factor."

"Knowledge" was added as a core factor in the competitiveness definition and was declared as a bridge between the competitive advantage theory of Porter (1990) and the comparative advantage theory of Krugman and Obstfeld, (2000).

Finally, Karl Aiginger (2018) while making policy recommendations for better industrial policy in the new globalized world, discusses the progress of competitiveness from a cost-oriented construct towards an outcome targeting area. A comparison is made by him between the low and high road strategies, where later is considered better. Accordingly, in the long run, the nations opting for the high road strategies of skills, innovation, productivity, and sophisticated markets are more prosperous than the countries choosing for the cost-focused strategies. Moreover, developed countries may benefit more from the high road strategies since the developing countries can easily get in the markets with their lower costs. However, these strategies need a different set of definitions and constructs.

To provide the background for a fresh definition of the term competitiveness in line with the high road strategies, Aiginger (2018) provides the evolution of competitiveness concept. He divides the constructs of competitiveness into three broad categories; among which the first one which is price competitiveness is based on the old version of the competitiveness concept. In this category initially cost alone and later cost along with productivity has a vital role for measuring competitiveness. The second category which is quality competitiveness includes structural elements based on the country shares and drivers of competitiveness based on different growth and welfare theories involving innovation, education, country, cluster, and social capital, etc. This is a very broad category and encompasses diverse drivers of competitiveness; lying in between the cost and outcome competitiveness. The third category is the outcome competitiveness which is further categorized under the old and new perspectives. In the old version, competitiveness is measured by outcome indicators like GDP and employment while in the new version 'the beyond GDP goals' like life expectancy, satisfaction, happiness, etc.; gain attention while measuring the construct of competitiveness. Finally, he concludes with his definition for the term competitiveness as a country's ability to provide beyond GDP goals to its native people, both in present as well as future. Aiginger (2018) declare this version of competitiveness as 'perspective of socio-ecological transition'.

**3.2 History of Competitiveness**

In the current global world, the development of countries is analyzed more based on their competitiveness. However, the concept in itself has rapidly evolved over centuries from the works of Adam Smith to Michael Porter and even further.

Economists categorize the development of competitiveness phenomenon into three major theoretical perspectives. These include the functional, the structural, and the behavioral perspectives (Melnyk, O & Yaskal, I. 2013; Mokiy, 2010). Ideas of Adam Smith and Michael Porter are considered representative of the behavioral perspective. According to this approach, competitiveness refers to recognizing the traits and approaches of business organizations' behaviors while facing contention for a more efficient demand in financial terms. Therefore the business organizations endeavor for producing competitiveness in the market, whereas the foundations of the perspective lie within the capital thoughts of consumers and the maximal fulfillment of their wants.

The second perspective as the structural perspective given by A. Cournot, F. Edzhwarth, E.Chemberlain, J.Robinson, is based upon the idea that a fair market structure is independent of particular business organizations and individuals. The market system and its work conditions are the only significant players. In such a case competition is collective production management, leading towards the relocation of resources among different branches, and the market is governed by multiple buyers of identical products.

Thirdly, Y. Schumpeter's ideas are categorized among the functional perspective. The functional perspective considers the economic development strategies of the business organizations to be based upon an innovative approach. The business organizations are required to minimize the production expenditure and fulfill rising consumer demand with brand new products.

Olczyk (2016) traces back the history of international competitiveness concept to the 1970s, considering it dominating research in the field of international economics. Accordingly, the Theory of competitive advantage, which is based on the idea of a country's specialization in the production of more efficiently produced products, has dominated the international trade theory before the 1970s (Krugman and Obstfeld, 2003). The Theory of absolute advantage given by Adam Smith in his book "Wealth of Nations" is seen as the earliest attempt, providing the basis for international trade and competitiveness (Salvatore, 2013; Krugman & Obstfeld 2003). While the Theory of Comparative Advantage given by David Ricardo in 1817 in his book Principals of Political Economy and Taxation (Salvatore, 2013) is seen as the most significant contribution towards the global trade theory and competitiveness. However, the comparative advantage theory became incapable of providing a basis for the huge two-way trade among industries of homogenous products after the Second World War. However, this mutual trade was possible due to economies of scale rather than the constant scale and perfect competition as assumed by Ricardo in his comparative advantage theory. (Smit 2010; Krugman 1990; Grubel & Lloyd 1975; Linder 1961; Vernon 1966)

Towards the end of the seventies, new trade theories based on monopolistic competition models were introduced (Krugman 1992). While both the comparative advantage theories and the new trade theories considered a specialization in the key role behind gains from trade, new trade theories were based on the assumption of economies of scale and not the comparative advantage

(Olczyk, 2016). Afterward, a shift towards the oligopolistic market structure was seen when for limiting the number of competitors, firm-level scale economies were considered adequate; and trade models were developed on the assumption of the oligopolistic market mechanism. Thus, trade of homogenous products among firms having internal scale economies as a significant feature; was considered bilaterally beneficial even in the absence of comparative advantage during the mid and late 1970s (Smith 2010; Krugman and Obstfeld 2003). But for all that, the production site was not identified under new trade theories (Smith, 2010).

Porter (1990) put an end to this shortcoming by introducing a new trade theory in the form of his diamond model and thus providing a comprehensive approach towards the competitive advantage of nations. Accordingly, the competitive advantage of countries is based upon four categories of country traits having the country demand conditions, country's factor conditions, country's supporting and related industries; and the structure, strategy, and rivalry faced by the country companies. Moreover, state policy and chance also aid the national competitiveness mechanism, however, these do not initiate the competitiveness process (Porter, 1990).

However, porter combines several contrasting theories into a single idea, as from the classical and neoclassical schools he picks the "factor conditions", the Rostow growth theory and product cycle theory provides links to the "demand conditions", derivation of related and supporting companies is done form the Marshall's industrial districts and the polarization theory, while Schumpeter provides the basis for "firm strategy, structure, and rivalry" (Olczyk, 2016). Though the diamond model is generally relevant in a country's international competitiveness, it has encountered some criticisms. Moreover, Porter has received criticism from both economists as well as management scholars. The economists criticized him for the generation of a weak relationship between the model and international trade theories in addition to the lack of a model's forecasting ability

(Smith, 2010). While the management experts criticized porter for ignoring the multinational exercises; and therefore Dunning in 1993 incorporated some additional variables like pro-competitiveness policies, government policies, and foreign direct investments (Dunning, 1993). After Dunning, some human variables that included professional, entrepreneur, bureaucrat, politician, and worker were also incorporated in the model (Cho et al. 2008).

Although porter's diamond model revolutionized international competitiveness knowledge, by adopting a composite technique towards evaluating this phenomenon; it unleashed a worldwide debate regarding indicators and determinants of competitiveness. Resultantly, the two most significant international competitiveness indices, namely the Global Competitiveness Index by World Economic Forum and IMD World Competitiveness Ranking by the International Institute for Management Development were launched, with their foundations from Porter's diamond model. However, the technique adopted by the world economic forum is in very close association with porter's diamond model since the WEF defines competitiveness as a composite index encompassing micro, macro, and policy-related variables (Schwab, 2014). While Porter also includes government capacity, microeconomic factors, and macroeconomic scenarios in defining international competitiveness.

Finally, in bibliometric analysis of international competitiveness, Mercedes, et. al. (2019) relate three most dominant theories to the concept of competitiveness. Accordingly, the first theory providing the basis for competitiveness concept is the famous David Ricardo Theory of Comparative Advantage given in the 19th century. The second theory closely associated with the competitiveness concept is Michael Porter's Theory of Competitive Advantage (1990), and the third one is Esser, et. al., Theory of Systematic Competitiveness (1996). However, Systematic Competitiveness theory (1996) considers the innovative environment generated by the working of state, social actors, business associations and companies, as a fundamental contributor towards raising productivity and increasing competitiveness of firms/companies. Whereas the determinants of systematic competitiveness are divided into four levels that are meta, macro, meso, and micro; which is a step beyond the already discussed micro and macro level combination in the existing competitive definitions and indices.

The history of competitiveness passed through different stages and interpretations during the last two centuries. However, Ricardo's comparative advantage and porter's diamond model are the evergreens and two most significant contributions which completely changed the dynamics of term competitiveness and ended up with the development of global competitiveness indicators.

**4.0 The Existing Competitiveness Indices / Measuring Competitiveness**

**4.1 National Competitiveness Index**

1. **The Climate Competitiveness Index**

The Climate Competitiveness Index; was published by the United Nations Environment Programme (UNEP) and the organization AccountAbillity in 2010. This index placed nations and regions based on their struggle towards a lower-carbon economy. It measured low carbon leadership based on two dimensions, climate performance, and climate accountability.

• Climate accountability refers to the procedure of formulation of a climate strategy by civil societies, businesses, and the governments, which covers all the basic challenges and opportunities; well thought, planned, designed, discussed, and modified accordingly, with the consent of all stakeholders. It is a sub-index of the main index, with four major elements represented by thirteen relevant indicators.

• Climate performance refers to the performance history of capability and action, shown by the civil societies, businesses, and the governments while deciding the inducements, formulating effective mechanisms, and lowering the carbon intensity during the low carbon commodities expansion. It is the second sub-index of the main index, also having four major elements represented by thirteen relevant indicators.

2. The Global Competitiveness Index

The Global Competitiveness Index (GCI) is the ranking of world countries in the Global Competitiveness Reports published annually since 2004 and was developed by Xavier Sala-i-Martin and Elsa V. Artadi. It is a composite index comprising twelve pillars of competitiveness. These include micro, macro, and policy-related variables as discussed in the above definitions of competitiveness. Concerning the time frame, only the short and medium-term dimensions of productivity drivers are considered in the GCI. Some drivers that are of vital importance from a sustainability viewpoint and in the longer term, are not included in the GCI. With such understanding, integration of drivers that were irresistible in the long run became mandatory. Resultantly, steps were taken which finally ended up with the introduction of The Sustainable Competitiveness Index.

3. The Sustainable Competitiveness Index

An initial version of the Sustainable Competitiveness Index (SCI), launched in The Global Competitiveness Report 2011-12 marks the fact that some sustainability components that are unimportant in the shorter term; can change the productivity in the longer term. Thus, the sustainability factor was added in competitiveness definition and was restated with the addition of "while ensuring the ability of future generations to meet their own needs" in the above-given GCI definition. Put differently, elements needed in making competitiveness sustainable concerning the environmental, economic, and social aspects, are addressed in the sustainability competitiveness index (SCI). The newly defined SCI is the integration of several long terms targeting concepts with the already short term targeting concepts of the Global Competitiveness Index. However, the phenomenon of climate change is not included in the index, even under the sustainability head. Though the significance of the phenomenon is discussed multiple times in different editions of the Global Competitiveness Reports.

4. **The Sustainability Adjusted Global Competitiveness Index**

The Sustainable competitiveness version of the Global Competitiveness Index (GCI) was later on modified into the sustainability adjusted GCI in the 2012-13 issue of the index and was under focus till 2013-14 reports. In this index, the GCI was environmentally and socially adjusted and a final Sustainability Adjusted Global Competitiveness Index was introduced. However, the phenomenon of climate change, despite being the most threatening global issue of the modern world, is again neglected in the index.

**4.2 Regional Competitiveness Index**

1. **The European Union regional competitiveness index**

The European Union Regional Competitiveness Index (RCI) is a competitiveness ranking of European countries by Paola Annoni & Lewis Dijkstra. The EU RCI defines regional competitiveness as "Regional competitiveness is the ability of a region to offer an attractive and sustainable environment for firms and residents to live and work". This index builds a multidimensional and commensurate scenario of the European Union regions' competitiveness. The sub-national level used by the RCI is considered a better approach in evaluating the performance and inequalities over time, than the national level. It is a composite index comprising eleven pillars of competitiveness that are based on the concepts of long-lasting development and productivity. The index measures the regional competitiveness of 268 regions across the 28 European Union member states.

**4.3 Sector Competitiveness Index**

1. **Competitive Industrial Performance Index**

The *Competitive Industrial Performance* Index is the ordering of world countries in *Competitive Industrial Performance* Reports published biannually and developed by Nelson Correa and Thomas Nice for the United Nations Industrial Development Organization. The index assesses 150 countries while exploring the nation's industrial sector contribution towards its development. To put it another way, the present index quantifies a country's industrial sector success in manufacturing goods; in trading its products within the local and global markets, and resultantly measures its contribution towards the structural changes and developmental progress of the country. It is a composite index comprising eight pillars of competitiveness. However, a new element was incorporated in the 2018 CIP index showing industrial production's impact on the environment. An adjusted CIP index was formulated in which higher carbon dioxide emitting countries descended to lower positions.

2. **The Global Manufacturing Competitiveness Index (GMCI)**

Global Manufacturing Competitiveness Index (GMCI), is a multi-year competitiveness ranking of world countries by Deloitte Touche Tohmatsu Limited (Deloitte Global) and the Council on Competitiveness. It has a total of three publications in 2010, 2013 and 2016. The index measures countries' capabilities in driving the structural changes in the manufacturing sector to anticipate the global economy. Towards an accurate quantification of the country's competitiveness, executives of manufacturing companies were requested to judge the overall competitiveness of the manufacturing sector in selected nations during the current date and the coming five years. The survey done from the executives was directly used to build GMCI, where the manufacturing attractiveness of each country was ranked. This composite index ranks about 40 countries of the world while comprising twelve drivers of competitiveness for global manufacturing.

### 3. The Travel & Tourism Competitiveness Index (TTCI)

This index is a global biannual ranking of countries by Lauren Uppink Calderwood and Maksim Soshkinby for the World Economic Forum. It defines travel and tourism competitiveness as "the set of factors and policies that enable the sustainable development of the Travel & Tourism (T&T) sector, which in turn, contributes to the development and competitiveness of a country." It is a composite index that ranks about 140 countries of the world while comprising fourteen pillars of competitiveness that are grouped into four sub-indexes.

**5.0 Climate Change and Competitiveness**

Porter and Linde (1995), declare that environment-competitiveness debate is misunderstood and needs to be reframed. They argue that the environment and industrial competitiveness can both progress if based on innovative solutions to the problems of industry-relevant to the environment. However, they do not give any consideration to the agricultural sector while discussing the environment-competitiveness relationship. Moreover, among most of the studies discussing climate change and competitiveness, the main emphasis has been on climate change-related policies and regulations affecting the competitiveness of different markets.

Bassi & Duffy (2016), while exploring the impact on competitiveness; made a contrast of the United Kingdom's Climate Change Policy with her competitor's policies and declare competitiveness unaltered. They find that with the promotion of special innovations and efficiency, climate change policies can have a beneficial impact on the UK's competitiveness. However, they declare competitiveness as a valid concern only for a limited number of sectors, which are more exposed to international trade and are also more dependent on energy use. However, this report does not mention the agricultural sector, or the impacts coming upon its competitiveness due to climate change.

Aldy & Pizer (2011), generate a clear-cut definition of the competitiveness impacts desirable for the climate change regulations using the domestic production and net import data. Further, making use of this definition they find the effects of energy prices on the carbon pricing policy. However, the focus of the study stays on the industrial sector and relevant climate change mitigation policies; without any idea of the country's agriculture sector.

In a report discussing the implications of the Kyoto Protocol on competitiveness and WTO's measures to neutralize the competitive losses, the definition of competitiveness based on the firms is used and the general concept of competitiveness of nations is ignored. The foundations for the definition of competitiveness are provided by Krugman (1994) and others, who argue that the concept of competitiveness has a similar interpretation as that of productivity when studied at the national level. They further argue that for the past half-century, national productivity and not the business terms of trade, has been the only concept used for quantifying the living standards of the masses. However, a more practical and genuine application of the competitiveness concept is to study it at the firm and sector level. Thus, with such backgrounds, competitiveness can easily be stated as an occupation of the market share, a position that is maintained in a dynamic contest among firms. Further, authors also discuss that measuring the competitiveness impact should be considered as only a single part of the greater effort of seeing the social welfare costs and benefits of action on climate change. (Cosbey and Tarasofsky, 2007)

In short, the changes in the market competitiveness of the agriculture sector due to the changing climate seems to be ignored.

## 6.0 Climate Change and Agriculture

The impact of changing climate has been studied by many researchers as can be seen from different studies (Kumar, & Sharma, 2013; Baig et al., 2011; Gornall, et al., 2010; Shakoor et al., 2011; Mendelson et al., 1992; Rehana et al., 2012; etc). The impacts on agricultural farm revenues, wheat production, and other major crops resulting from climate change were studied in the above studies by using different climate variables like rainfall, temperature, etc. Most of these studies concluded that climate change had a measurable impact on farm productivity, costs, and revenue. Studies are also available on effects upon food security, adaptation, and mitigation policies due to climate change; like Wheeler & Braun (2013). However, researches finding climate change impacts agricultural market competitiveness was scarce. Thus, the impact of changing climate on the market competitiveness of the agricultural sector seems to be a grey area that needs to be explored to find the existent relationship.

## 6.1 Climate Change and Agricultural Productivity

The impact of changing climate on agriculture productivity is already studied from multiple dimensions. The estimations in the study of Janjua et al., (2010) reveals the absence of any impact of global climate change on Pakistan's wheat productivity. While Raza (2015), discovered a significant impact on cotton production due to the temperature and precipitation changes. Further, Zhu et al., (2011) declared a negative effect of the changing climate on crop production due to the varying supply of irrigational water resulting from climate changes. However, the impacts of changing climate on agricultural production differ depending upon the geographic location; age, soil, and kind of trees, carbon dioxide, and nitrogen fertilization and interactions among any of these factors. (Girardin et al., 2008; LeBauer and Treseder, 2008; McMillan et al., 2008; Ollinger et al., 2008; Phillips et al., 2008; Reich and Oleksyn, 2008; Saigusa et al., 2008; Clark et al., 2003). Among other factors, Pakistan's agriculture is more susceptible to climate change due to the country's geographical location and the usual warm climate.

The major approaches used in determining the climate impacts on agriculture include a Ricardian Approach, a crop modeling approach, and a panel data approach. The Ricardian model was greatly promoted by the works of Robert Mendelsohn and his fellow workers (Mendelsohn et al., 1994; 2001; Mendelsohn and Dinar, 2009; Sanghi and Mendelsohn, 2008). Further, the same model was used by Shakoor et al (2011); and Kumar and Parikh (2001); in their studies.

**7.0 Agriculture and Competitiveness**

Gardiner et al. (2004) state that only nations approaching high productivity indexes and record-high employment rates can be competitive. While Laureti and Viviani (2011), established that firm competitiveness considerably gets effected by productivity. These outcomes endorse Porter's theory suggesting that firm competitiveness and productivity are analytically linked up, making themselves helpful in improved decision making for increasing competitiveness at the firm level. While several scholars stress that national productivity is the sole representative phenomenon for competitiveness measurement Porter (1990). Just like the firms, the farm (agricultural) productivity serves as an eminent determinant of the agriculture sector market competitiveness.

Furthermore, agricultural costs are also an important determinant of agricultural market competitiveness. The studies of Gallagher et al., (2006), Thorne (2005), Mulder et al., (2004), and Bureau & Butault (1992); made the use of agricultural costs for finding competitiveness as domestic wages to labor productivity ratio, by comparing unit labor cost, calculating costs of production as an average over the period, and by measuring different cost indicators like total costs as a percentage of the value of total output. Some studies like Sharples (1990), included the marketing costs along with the production costs for finding agricultural market competitiveness.

**8.0 Climate Change and Agricultural Market Competitiveness**

Climate change is predicted to affect European agriculture and is foreseen as a threat to the emanating and recurring feed and food problems. The indirect effect of such hazards might include the occurrence of mycotoxins happening as a result of the geographical modifications in the arrangement of vital cereal cropping systems, thus impacting crop rotations. With such backgrounds, Elsgaard et al., (2012) study the climate change impact upon the cropping shares of wheat, maize, and oat, on a 50 km square grid across Europe. This paper provides projections for 2040, of the model-based approximations of changing cropping shares resultant from the different temperature and precipitation.

Frank et al., (2014) provide a consolidated analysis of both demand and supply-side impacts of climate change on the agricultural sector of Europe. They use partial equilibrium models for the purpose with a focus on Europe and conclude that agricultural demand, supply, and even the producer prices will be remarkably influenced by climate change. The demand side impacts can be minimized through adaptation mechanisms like reapportionment of resources or intensification to hold the preliminary climate shocks. They effectuated an initial basic framework by connecting CAPRI and GLOBIUM-EU, where they run two climate change structures. Changing climate instigated productivity disturbances contrasted to an initial basic framework without climate change and showed a worldwide cutback in the crop supply along with associated price hikes.

Through feeding costs, the livestock sector is affected indirectly as well from the changes in crop supply leading to higher prices. However, the European consumers are comparatively not much impacted by climate change about diminishing calorie consumption, while other than Europe, people face elevated foodstuff prices caused by climate change. Moreover, the general response patterns reflected in the global models clarified that adjustments on the supply side are stronger compared to rather inelastic behavior on the demand side, and prices are the most sensitive parameter affected by climate change (Nelson et al., 2013). Despite these shortcomings, the results find out a negative effect of climate change upon European consumers and producers, through falling productivity and rising prices. However, the agricultural market competitiveness impact of climate change which is in close relation to the price changes is not considered in the paper.

The agricultural setup of the European Union was considerably transformed by the cycle of the Common Agricultural Policy reforms from 1992 to 2013. The operation consisted of a continuous but limited and irregular liberalization of European agriculture. This was due to more focus on destroying the intervening apparatus rather than reorganizing, contemporizing, and modifying towards a greater competitive domain. Thus, with such a background, the states were incentivized to formulate policies to accelerate individual competitiveness and innovative capacity under the setup of Common Agricultural Policy (2014-2020). Under the same assertion, Moralejo & Sanchís (2017) report the basic quantitative and qualitative hidden facts of regional agricultural policy prevalent in Spain. However, apart from reporting the hidden facts, the researchers also mark the need for future research "to look for viable strategies to boost competitiveness and innovation in the agriculture of the different Autonomous Communities (ACs)".

Carraresi & Banterle., (2013) makes a comparison of the agricultural sector with the food industry while assessing the sector level competitive performance of the European countries within the

European Union market. The paper aimed to find out the effects coming upon the competitiveness of the EU from the first phase of European Union enlargement and the second phase of economic crisis. However, the competitive performance of the countries was calculated by making use of the different trade indices; such as Export and Import Market Share, Revealed Comparative Advantage, Net Export Index, and Vollrath indices. The findings revealed insignificant changes in the food sector and agricultural competitive performance for the past fifteen years.

Hermans et al., (2010) made an exploratory study on the crop production of Europe in a liberal world, focusing on climate change and competitiveness dimensions. To evaluate the European agricultural competitiveness, demography, technology, climate, and international demand for agricultural commodities were used. The indicators like economic strength of farms in a region along with the population pressure on agricultural land were used to assess the sustainability of agricultural production in a liberalized market scenario. Their findings revealed that notwithstanding a few unfavorable impacts from climate change, the future need for food can be fulfilled with a smaller land area due to the technological progress in the sector.

However, the study is exploring the future agriculture market scenario in case of complete liberalization of the same markets, by eliminating the trade restrictions and subsidies; and is not focused on competitiveness alone. Moreover, the study is focused on the international competition and trade aspect of competitiveness and ignores the sustainability and productivity aspects of competitiveness. The study also does not focus on the unit of analysis used for finding out the agricultural market competitiveness. Furthermore, the proxies used for finding competitiveness are not sufficiently supported in previous literature. Finally, the study is focused on the international competitiveness of the agricultural market and there is no consideration of the regional and sector competitiveness. Exploratory research is needed to find out the appropriate proxies for the conduction of the study.

Another study by Costinot (2016) has used the already existing GAEZ models, to sum up, the individual impacts of the changing climate on the produce of different countries. The study makes use of micro-level studies to provide a macro-level knowledge regarding how climate change affects the agricultural markets. Moreover, Costinot (2016) relates the impacts to comparative advantage and global dispersion of the crops through trade, while considering the productivity measures provided by agronomists, before and after climate change as his main variable of interest. He comes up with the finding that adjustments in production by using the comparative advantage phenomenon can alleviate the harmful outcomes from the changing climate.

However, competitiveness is a relative term and can be quantified in multiple ways, and productivity cannot be the only measure. Market share, total revenue, and total sales, etc., can also be used as proxies. As productivity can only provide knowledge regarding the output but the loss of product during transportation and dispersion in the market due to inappropriate temperature etc., can be the cause of huge losses in product and thus the competitiveness of crops. Moreover, the paper focuses on the trade aspect and comparative advantage. While competitiveness is not necessarily a trading concept, it can also express the sustainability of crops and agricultural markets. Therefore, for quantification of climate change impacts at the micro-level, the data set is based on the productivity measures provided by agronomists, both before and after climate change. However, the agronomists' estimates might be biased due to insufficient farm records, memory loss in old agronomists, frequent fragmentation of land, etc. Since the study is based on secondary data and the proxies used are not sufficiently supported in previous literature; exploratory research is needed to find out the appropriate proxies for the conduction of the study.

**9.0 Proposed Strategies for Agricultural Competitiveness Improvement**

Attila and Suresh (2016), proposed nine strategies for improving agricultural competitiveness based on the nine determinants identified by them. They considered the world's economic forum's framework of global competitiveness index and its twelve determinants of global competitiveness. Further, from the previous studies discussing the factors impacting the competitiveness of the agriculture sector of countries, key factors responsible for competitiveness in the agricultural sector were also recognized. Finally, with the twelve global competitiveness determinants and the global agricultural competitiveness determinants, they figured out nine key factors responsible for competitiveness in the agricultural sector, and based on these determinants; they proposed nine strategies for improving global agricultural competitiveness. Moreover, they declare that a merger general and agriculture specific determinants of competitiveness; is the only possible way for measuring global agricultural competitiveness.

The strategies proposed for increasing agricultural competitiveness

1. Efficient organizations
2. Sustainable handling if available natural resource
3. Real/ Physical infrastructure investment
4. Innovative and advanced technology adoption
5. Favorable environment creation
6. Advanced risk management mechanism of agriculture
7. Health and education spending

8. Effective land markets

9. Better market access

**10. Conclusion**

The development of the agricultural sector is declared as one amongst the most robust elements to eliminate extreme poverty, enhance mutual welfare, and feed an estimated number of 9.7 billion people by 2050. This sector's growth in comparison to other sectors is considered from double up to quadruple times more successful in improving incomes of the lowest ebb of a society. The 2016 World Bank analyses revealed 65% of the poverty-stricken adults earning their livelihood through this sector. Moreover, one-third of the world's gross domestic product was generated through the agriculture sector in 2014. However, all this food security, poverty alleviation, and economic growth generated from the agriculture sector are highly at risk. Most of the crop yields are already getting impacts from the changing climate, particularly damaging the most food-insecure regions. Thus, in 2020, the huge impacts from changing climate, warfare, pests, and spreading infectious diseases are anticipated to jeopardize the food production, damaging the supply chains and overstretching/pressurizing people's capabilities to acquire nutritive foods at economical/reasonable prices. (World Bank, 2020)

On the other hand, the development of countries is analyzed more based on their competitiveness in the current global world. Competitiveness in agricultural and food markets is gaining importance rapidly not just in developing countries, but the world over. It can affect the working of those markets in many ways and becomes a tricky question. Moreover, the absence of competition brings a direct impact on producers and consumers in the agriculture and food markets. These impacts include price instabilities, price transmission mechanisms, availability and access to the products, etc. Besides, the absence of healthy competitive forces in these markets may lead to failure of the government policies targeting the same markets. However, the impacts coming upon farmers might differ from the impacts coming upon the producers depending on the food security measurement tools used. (The State of Agricultural Commodity Market in Depth 2015-16)

Countries like Australia with highly developed agriculture, allocate billions of dollars for promotion of competitiveness in agriculture. The Agricultural Competitiveness White Paper (December, 2017) is an Australian Government's plan to grow agriculture and the government is investing $4 billion in their farmers under the plan. According to the white paper, "Stronger farmers mean a stronger Australian economy".

Thus, improvement of competitiveness in agriculture can help the farmers to get fairer farm gate returns, help to improve and secure the farming infrastructure, bring in better preparedness for calamities and climatic risks management, and finally flourish the country's international trade. Moreover, it can also help in increasing consumer welfare in the form of competitive prices and improved quality as an outcome of a competitive market.

Different competitiveness indices are used for measuring different sectors' progress, development, and stability. However, we cannot find any index for measuring agricultural market competitiveness. Though agricultural market competitiveness is measured by making use of individual measures but no composite index is constructed for the purpose. While the previous/above discussion regarding definition, history, and measurement of competitiveness manifests that a composite index and not a single measure can best measures the competitiveness of a sector. Thus, the construction of an agricultural market competitiveness index is proposed to be an appropriate measure for analyzing the problem empirically. Moreover, Schwab (2012) declares climate change to be a very important phenomenon, which needs to be incorporated into the existing competitiveness indices. Also, climate change has significant impacts on the agricultural sector, and therefore to construct an agricultural market competitiveness index without including the climate change factor will make the index a biased approach. Therefore, we will incorporate the phenomenon of climate change in our proposed agricultural market competitiveness index. This will not only help in enhancing the potential of agricultural markets for optimizing the agricultural sectors' competitiveness but will also help in designing the agricultural market-friendly climate policies and ensure sustainable growth in the sector.